\documentclass{kluwer}

\usepackage{graphicx}

\begin{document}
\begin{article}
\begin{opening}

\title{The application of {\it \`a Trous} wave filtering and Monte 
Carlo analysis on SECIS 2001 solar eclipse observations}

\author{A.C. \surname{Katsiyannis}\email{T.Katsiyannis@oma.be}}
\institute{Royal Observatory of Belgium, Avenue Circulaire -3- Ringlaan, 
1180, Brussels, Belgium}
\author{F. \surname{Murtagh}} 
\institute{Department of Computer Science, Royal Holloway, University of 
London, Egham, Surrey TW20 0EX, UK. \\
Observatoire Astronomique de Strasbourg, 11 rue de l'Universit\'e, 67000 
Strasbourg, France.}

\runningtitle{{\it \`A Trous} filtering and Monte Carlo analysis on 
SECIS data}
\runningauthor{Katsiyannis \& Murtagh}

\begin{abstract} 
8000 images of the Solar corona were captured during the June 2001
total Solar eclipse. New software for the alignment of the images and
an automated technique for detecting intensity oscillations using
multi scale wavelet analysis were developed. Large areas of the images
covered by the Moon and the upper corona were scanned for oscillations
and the statistical properties of the atmospheric effects were
determined. The {\it \`a Trous} wavelet transform was used for noise
reduction and Monte Carlo analysis as a significance test of the
detections. The effectiveness of those techniques is discussed in
detail.
\end{abstract}

\keywords{Sun, Solar corona, Image processing, Instrumentation}

\end{opening}

\section{Introduction}

A number of theoretical predictions exist for the propagation of
magnetohydrodynamic (MHD) waves associated with coronal
loops. \inlinecite{Roberts84}\ studied the various types of
propagation through a low-$\beta$\ plasma using reasonable
approximations for the conditions inside coronal loops. Their work was
followed by a large number of authors since then (for example see one
of the many review papers by \opencite{Nakariakov03}) and was recently
confirmed and refined by applying numerical modelling
\inlinecite{Nakariakov04}.

Many attempts to observe propagations in the coronal loops have been
made since the first theoretical predictions. One of the most
challenging types of oscillations to observe are the fast sausage mode
MHD that have expected periodicities below 1$min$\ (see
\opencite{Aschwanden04}\ for a detailed review). Radio, optical and 
X-rays observations have been used to detect such waves with limited
success. In this paper we will present the application of image
processing techniques as a way to enhance optical observations made by
the Solar Eclipse Coronal Imaging System (SECIS) project during the
June 2001 total solar eclipse in Zambia. A detailed description of the
instrument can be found in \inlinecite{Phillips00}.

Starting with \inlinecite{Koutchmy83}\ a number of authors have
published possible detections of oscillations with periodicities below
10$sec$. \inlinecite{Pasachoff84,Pasachoff87}\ have reported possible
detections of optical intensity oscillations with periods in the range
of 0.5--4$sec$, while more recently
\inlinecite{Williams01,Williams02,Katsiyannis03}\ (hereafter W01, W02 
and K03 respectively) provided strong indications of oscillations with
periodicities $\approx$6$sec$ while reporting on optical SECIS August
1999 total solar eclipse observations.

Continuing the work published for the SECIS 1999 observations, we have
analysed observations made during the June 2001 total solar eclipse in
Lusaka, Zambia. Based on experience from the analysis of the 1999 data
set, a number of numerical techniques were used in order to improve
the signal-to-noise (S/N) ratio and establish an objective, numerical
criterion for the identification of the corona intensity oscillations
over any statistical effects. A brief description of the observations
and data analysis is presented here with more emphasis given to the
advanced mathematic techniques used in an effort to improve the S/N
ratio and determine the \lq \lq real'' detections of corona loop waves
over the influences of noise in the data set.

\section{Observations \& Data Reduction}

A detailed description of the SECIS instrument, as used for the 1999
observations is provided by \inlinecite{Phillips00}, while a
discussion of the improvements made for the 2001 observations can be
found in \inlinecite{Katsiyannis04}\ (hereafter K04). The observations
taken by SECIS in 2001 and their data reduction will not be presented
in detail in this paper as they are the subject of K04. However, a
brief description of the data set is to follow as needed for the
presentation of the image processing techniques reported here.

8000 Fe~{\sc xiv} images of 512$\times$512 pixel$^{2}$ with a
resolution of $\approx$~4~arcsec~pixel$^{-1}$ were taken during the
$\approx$~3.5~$min$ of totality. Although the observing field was
large enough to include the whole Moon disk and the lower part of the
corona, we chose only to observe the North-East limb. This decision is
in line with the 1999 observations and was taken to avoid edge effects
of the CCD and optics as well as to include important parts of the
outer corona.

A brief description of the data reduction of the eclipse 2001
observations is included here for the purpose of describing the data
that were used for the application of the {\it \`a Trous} wavelet
transform and Monte Carlo analysis. The images taken during the 2001
observations were reduced by using dark and flat-field frames , for
current subtraction and flat field correction. The 8000 images were
then automatically co-aligned using the edge of the Moon as a
reference point for a first order alignment. For this first alignment
the moon was effectively considered stationary during the duration of
the eclipse. A more accurate alignment was subsequently achieved by
using a clear feature from an area of the lower corona as
reference. This second alignment corrected for the motion of the moon
in respect to the solar corona during totality. K04 provides a
detailed discussion on the alignment technique used and its various
steps.

After co-alignment the 8000 images of the observations form a three
dimensional data array. The basic technique used for the detection of
intensity oscillations throughout the SECIS project is the continuous
wavelet transformation of the time series that corresponds to each of
pixels of the aligned images. Details on the transformation function
and its implementation in time series can be found in
\inlinecite{Torrence98}, while examples of the application of this 
analysis to SECIS data can be found in a number of publications (e.g.\
W01, W02, K03 and K04). Additionally K03 explicitly mentions a number
of criteria used for a wavelet detection to be considered as a solar
intensity oscillation (as oppose to detections created by noise). To
test the performance of K03's criteria, K04 used automated software to
scanned large areas of the image covered by the Moon and upper corona
and confirmed their effectiveness.

\section{Noise Reduction using the {it \`A Trous} wavelet algorithm}

\subsection{S/N ratio limitations on SECIS eclipse data}

On of the most significant limitations of the SECIS eclipse
observations is the low S/N ratio. Although the observations were
taken using a broad Fe~{\sc xiv} filter and the solar corona is known
to be bright in Fe~{\sc xiv} emission, there are three major factors
that severely limit the S/N ratio achieved by SECIS.  For the purposes
of this paper we will only emphasise the S/N ratio limiting factors:

\begin{enumerate}

\item The prime mirror of the SECIS telescope has a 200 $mm$ diameter 
and a focal length of f/10. This is because the instrument was
designed mainly with observing solar eclipses in mind and has to be
lightweight and easy to travel.

\item The CCD cameras took observations at a ratio of $\approx$40 
frames per second. This has the obvious disadvantages of a very short
exposure time and a very fast readout speed. With current technology
such fast CCD readout speeds increase the readout noise drastically.

\item As the purpose of the SECIS project is to detect high frequency 
intensity oscillations, the atmospheric effects became
significant. This is because the earth's atmosphere itself is known to
oscillate in these frequencies and that is causing non-Gaussian noise
on the data set.

\end{enumerate}

\subsection{Advantages of using the {\it \`a Trous} wavelet transform 
in our data set.}

Having the above limitations in mind, the {\it \`a Trous} filtering
was investigated as a means of noise reduction because the algorithm
has a number of advantages \inlinecite{Starck02}.

\begin{itemize}

\item The computational requirements are within acceptable 
levels. This is particularly important as both of the SECIS eclipse
observations will have a size of more than 8 $GB$.

\item The reconstruction algorithm is trivial. This is important as it
makes the reconstruction of the time series more accurate.

\item The transform is known for every sample of the time series of 
every pixel. This is important to this project as the exact moment an
oscillation starts and ends on a given pixel can be very
significant. W02 made some interesting measurement of the propagation
speed of a travelling wave on SECIS 1999 total solar eclipse
observations by determining the exact time the oscillation arrived at
any given pixel.

\item The transform is clearly evolving through the different scale in 
a predictable mater. This makes easier to choose the right scale for
the filtering of a certain data set.

\item The transform is isotropic. As the SECIS data are also isotropic 
(specially in the time domain), any filtering used should also be
isotropic to avoid an artifacts being introduced.

\end{itemize}

The {\it \`a Trous} filtering is a relatively recent (some of
the first application were described by \opencite{Holschneider}), s
sophisticated, highly tunable and complicated multiresolution
algorithm. Due to its complication the exact description of the
algorithm is outside the purposes of this paper. More information on
the algorithm, its various parameters, advantages and disadvantages of
the various procedures that can be used in conjunction with the {\it
\`a Trous} wavelet transform and examples of its application on
astrophysical data can be found in
\inlinecite{Starck02}.

\subsection{The application of {\it \`a Trous} noise reduction 
algorithm to SECIS data.}

Each pixel of the three-dimensional data array of the reduced SECIS
2001 observations was treated as a independent time series and was
transformed to a number of coefficients on a multi-scale domain using
the {\it \`a Trous} wavelet transform. $B_{3}$ splines were used
for correlation and the noise was assumed to be Gaussian. The sigma of
the noise in the coefficients was determined automatically and
multiresolution hard K-Sigma thresholding was used to remove the
coefficients that were found to be noise. The time series was then
reconstructed using the coefficients corrected for noise.

\begin{figure}
\vspace{1pc}
\includegraphics[bb=0 190 504 510, width=11cm]{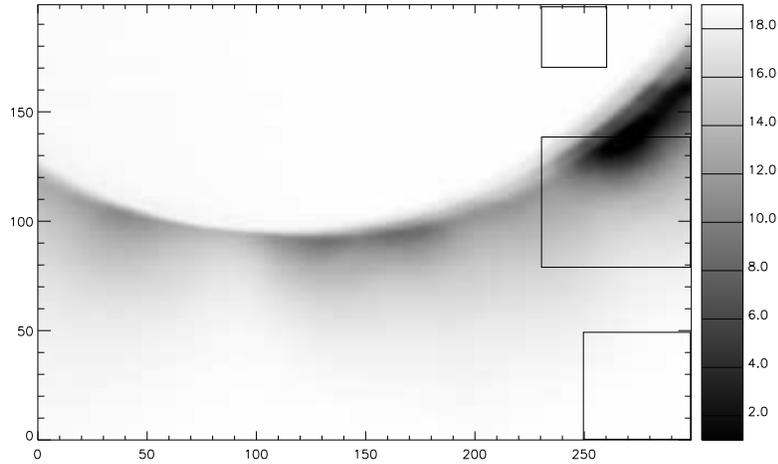}
\caption[]{The 2001 total solar eclipse as observed by SECIS. This 
image is produced by averaging over the time axis the
three-dimensional aligned array. After alignment the Sun remains in
the same pixel area and the Moon crescent moves in respect to the
solar corona. Areas of the Moon, outer corona and lower corona close
to NOAA Active Region 9513 (AR 9513) are highlighted. These three were
the areas used to test the effectiveness of the {\it \`a Trous} noise
filtering and the Monte Carlo randomisation analysis and were also the
exact areas also used by K04.}
\label{areas}
\end{figure}

To test the effectiveness of the {\it \`a Trous} wavelet
transformation in noise reduction the same three parts of the data set
that were chosen by K04 were used again. Figure \ref{areas} contains
the time-average image of the aligned SECIS 2001 total eclipse
observations. After alignment the three-dimensional data array was
averaged over the time axis resulting to a two-dimensional image. The
edges of the image (one hundred pixels of the left, right and bottom
of the image and two hundred pixels from the top) were discarded as
they suffer from edge effects of the CCD. Highlighted are three areas
of the data set that were used to test the effectiveness of the {\it
\`a Trous} filtering algorithm and the Monte Carlo randomisation
test. These are areas of the image covered by the Moon's disk, the
outer corona and lower corona. Those areas were chosen to be in the
proximity of AR 9513 as this is potentially the most interesting (from
the coronal loop oscillations point of view) area of the 2001
observations. The proximity of the test areas to the area were more
detections of coronal waves are expected is important as it produces
more reliable statistics as there are no effects from large-scale
variation in CCD sensitivity or deferent atmospheric conditions.

\begin{figure}
\includegraphics[bb=0 100 504 610, width=5cm]{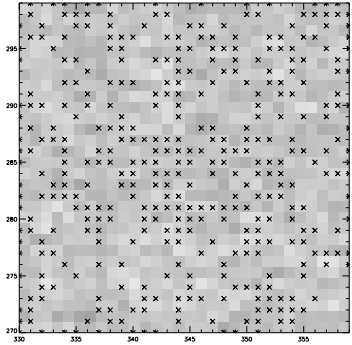}
\caption[]{A 30$\times$30 pixels Moon area of the 2001 observations. 
All pixels in that area had their Gaussian noise removed by applying
the {\it \`a Trous} wavelet transformation. Marked with \lq x'
are the 374 detections made by the automated software of K04 after the
removal of Gaussian noise.}
\label{moon}

\end{figure}
\begin{figure}

\vspace{1pc}
\includegraphics[bb=0 140 504 580, width=8cm]{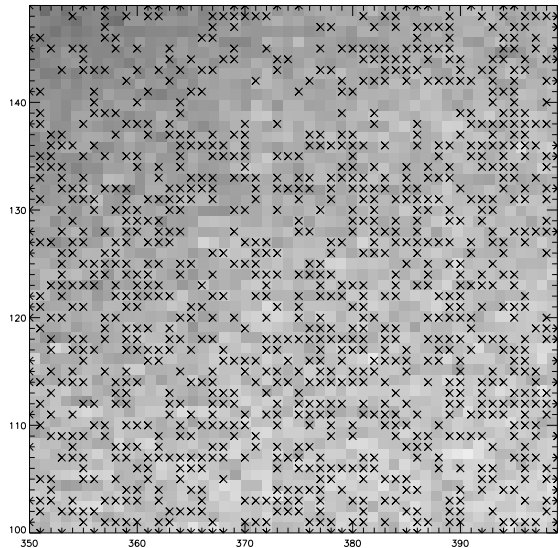}
\caption[]{A 50$\times$50 pixels area of the outer corona. The signal 
of these are was filtered using the {\it \`a Trous} filtering
algorithm and the automated technique of K04 was used for the
detection of intensity oscillations. 1054 pixels found to oscillate in
intensity.}
\label{upper}

\end{figure}
\begin{figure}

\vspace{1pc}
\includegraphics[bb=0 150 504 570, width=11cm]{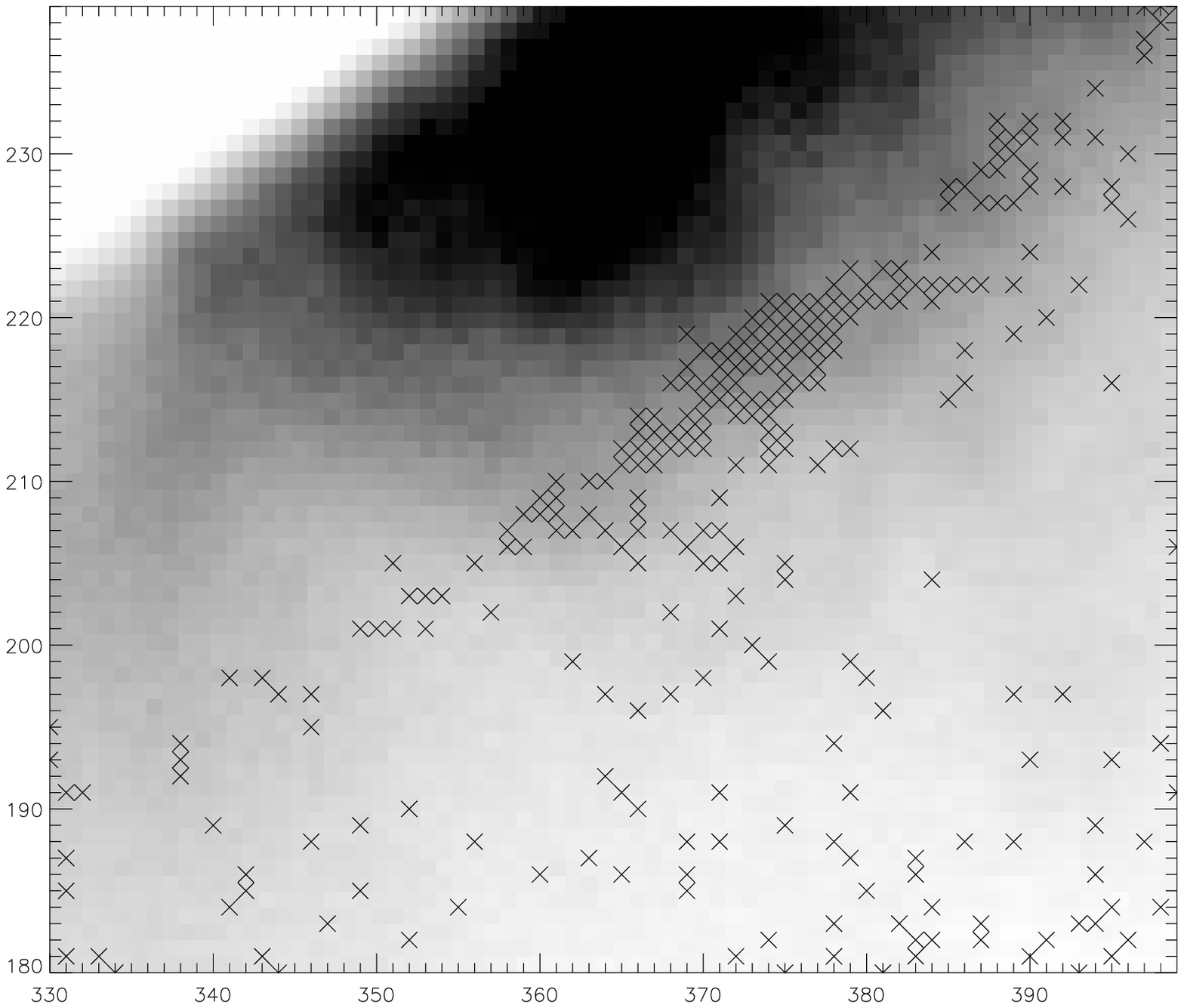}
\caption[]{A 70$\times$60 area of the lower corona and the Moon limb. 
The area was treated with the same algorithm as Figures \ref{moon} and
\ref{upper} and 276 detections were found.}
\label{lower}

\end{figure}

The same automated detection algorithm was used as in K04 and the
results for the periodicity range of 7--8 $sec$ are displayed in
Figures \ref{moon}, \ref{upper} and \ref{lower}. Figure \ref{moon}
contains 374 pixels that have false (i.e.\ not caused by the solar
corona) detections of oscillations out of the 900 pixels of the
sample. Before the {\it \`a Trous} noise filtering the same area
contained 5 oscillating pixels. Figure \ref{upper} contains 1054
oscillating pixels out of 2500 while before the application of the
{\it \`a Trous} wavelet transformation algorithm there were 11. On
figure \ref{lower} we have detections on 276 pixels out of the 4200
pixels of the sample while K04 found 84 (out of which 66 were
concentrated in a very compact area in the middle of the image).

The difference in number of detections before and after the is
significant. The number of detections before and after the {\it \`a
Trous} wavelet transformation increased by a factor of 75, 96
and 3.3 for the Moon, outer corona and lower corona areas
respectively. What is also important is that the increase is not the
same for the three areas. While the increase in number of detections
is similar for the Moon and outer corona areas, it is by far smaller
for the lower corona.

\begin{figure}
\vspace{1pc}
\includegraphics[bb=19 19 576 721, width=11cm]{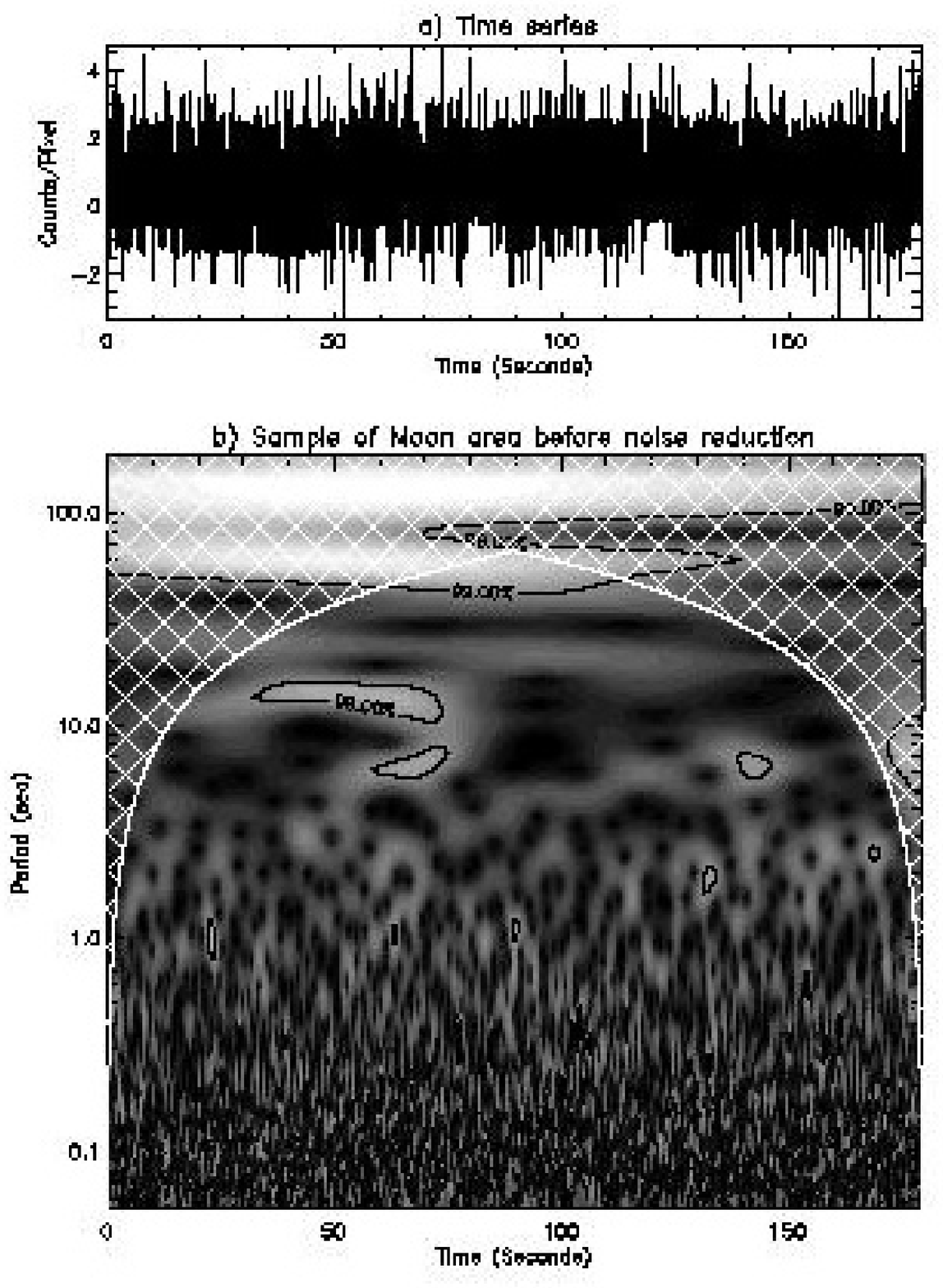}
\caption[]{Wavelet analysis of pixel x=344, y=281 of the Moon area. 
Panel (a) contains the time series while panel (b) the wavelet
transformation. This is the un-filtered time series and there are no
detections of oscillations that satisfied the criteria established by
K03.}
\label{wavelet_moon_noisy}
\end{figure}

\begin{figure}
\vspace{1pc}
\includegraphics[bb=19 20 575 735, width=11cm]{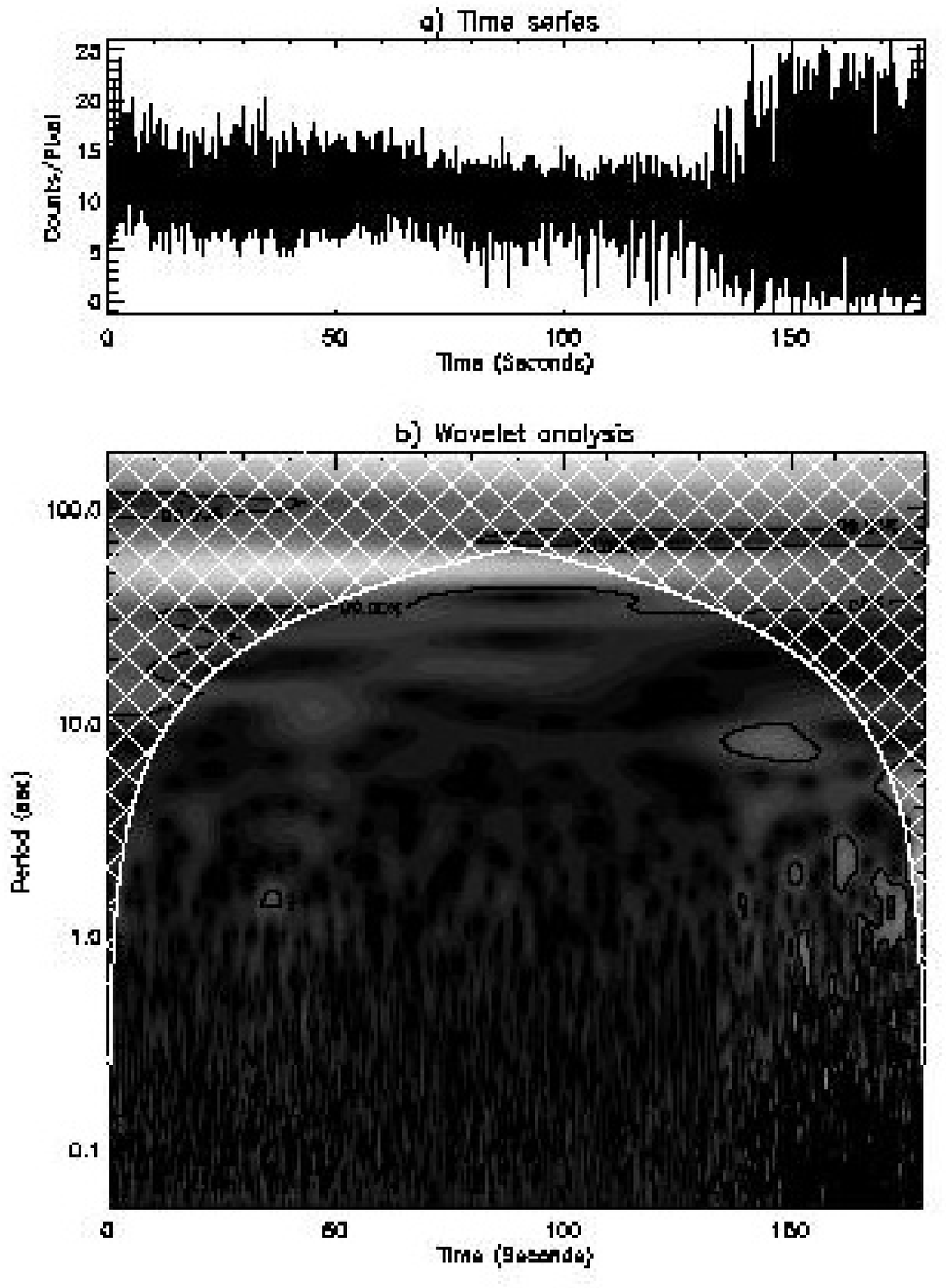}
\caption[]{Wavelet analysis of pixel x=374, y=221 of the lower coronal
area. Similarly to Figure \ref{wavelet_moon_noisy}, this is the
un-filtered time series and there are no detections of oscillations.}
\label{wavelet_lower_noisy}
\end{figure}

Although it is not surprising that the {\it \`a Trous} filtering
effects the areas with high S/N ratio less than those with very low
(or zero) S/N, it might appear strange that the filtering causes the
areas with very low S/N ratio to be detected as oscillating. To
examine this difference in some detail the wavelet transformation of
two pixels, one from the Moon area and another from the lower corona
are included as figures \ref{wavelet_moon_noisy}\ and
\ref{wavelet_lower_noisy}. Both points were not found to oscillate
before the filtering but only after. By examining the time series on
Panel (a) of Figures \ref{wavelet_moon_noisy}\ and 
\ref{wavelet_lower_noisy}, it is obvious that they are both very 
noisy, although in the case of Figure \ref{wavelet_lower_noisy}\ there
is an underling longer timescale variation, while in Figure
\ref{wavelet_moon_noisy}\ the signal oscillates around an average
value. These differences are also apparent in the panel (b) of the two
figures were the wavelet transformation in Figure
\ref{wavelet_moon_noisy}\ has a lot of high values in very low
periodicities (since those are far more effected by non-systematic
noise), while Figure
\ref{wavelet_lower_noisy}\ having higher S/N ratio is less effected by
noise therefor there are less high values in the wavelet
transformation even in low periodicities. On high periodicities,
although there is an area of interest in both figures, there is
nothing that satisfies the criteria established by K03.

\begin{figure}
\vspace{1pc}
\includegraphics[bb=19 19 576 769, width=11cm]{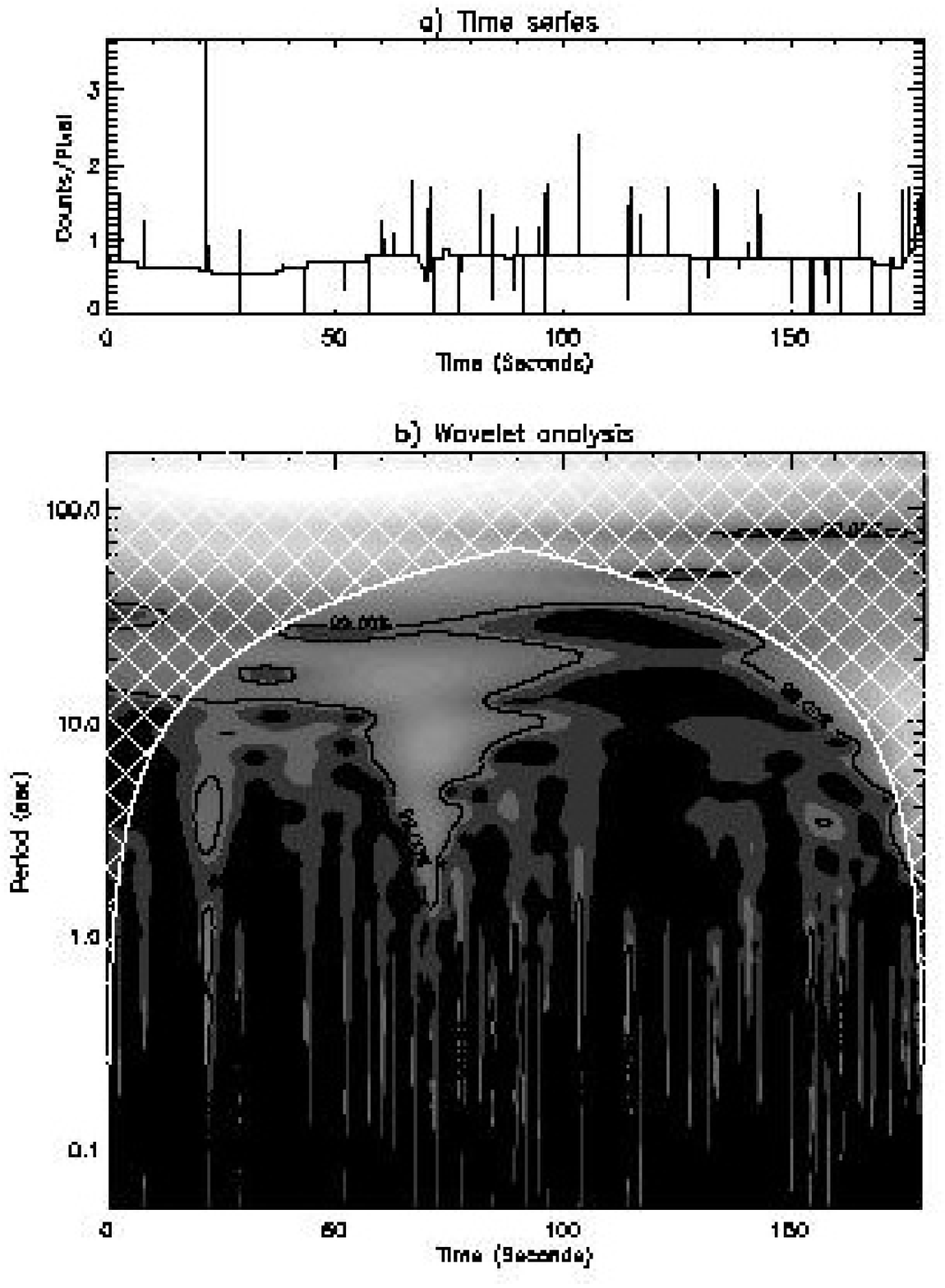}
\caption[]{Wavelet analysis of pixel x=344, y=281 of the Moon area 
after the time series was filtered through the {\it \`a Trous} wavelet
transformation. A number of oscillation can now be found.}
\label{wavelet_moon_filtered}
\end{figure}

After filtering with the {\it \`a Trous} wavelet transformation
algorithm the same two points were analysed using wavelets. Figure
\ref{wavelet_moon_filtered}\ contains the time series produced and and
wavelet transformation that corresponds to point of the Moon area that
we analysed in Figure \ref{wavelet_moon_noisy}. All the jittering in
the time series has disappeared and only some small picks and a small
long-term variation have remained. The wavelet transformation
corresponds well to what appears on the time series, producing
very low values on the very high frequencies (as the Gaussian
noise effects the high frequencies more), some short-lived high
values on the high frequencies (that correspond to the high
picks of the time series) and a number of detections in lower
frequencies (that correspond to the long-term variation).  As
expected, the {\it \`a Trous} wavelet transformation filtering
was very effective on removing the Gaussian noise (which is why there
are no oscillation in very low periodicities) and the detections on
the higher periodicities should be attributed to another factor. Since
by definition the area of the images covered by the Moon has not
direct light from the lower corona, another source of light should be
considered. As it is known that in total solar eclipses the sky is not
completely dark (the sky is much brighter that during night even to
the naked eye), the long-term variations in the time series and the
resulting detections should be attributed to the scattered light and
atmospheric affects that produces variations in brightness. Although
those existed previously in the un-filtered data, there were small and
effect by Gaussian noise, therefor they did not produce enough power
to become valid detections.

\begin{figure}
\vspace{1pc}
\includegraphics[bb=19 19 575 748, width=11cm]{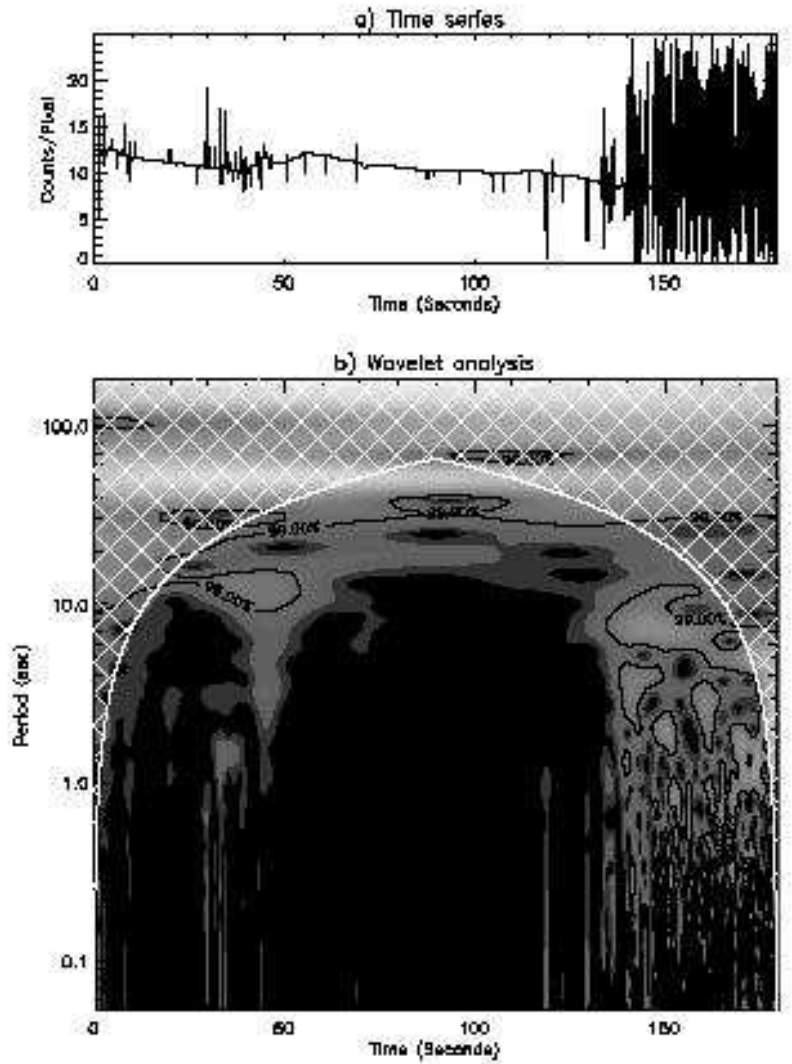}
\caption[]{Wavelet analysis of pixel x=374, y=221 of the lower coronal
area after the time series was filtered through the {\it \`a Trous}
transformation. As with Figure \ref{wavelet_moon_filtered}, a number
of oscillations can be found.}
\label{wavelet_lower_filtered}
\end{figure}

Figure \ref{wavelet_lower_filtered}\ contains the time series and
wavelet transform of the point of lower corona analysed in Figure 
\ref{wavelet_lower_noisy}. The effects of the {\it \`a Trous} 
wavelet filtering here are different to those on the previous
time series. Although most of the jittering has disappeared up to the
130$^{th}$ $sec$, the variation that existed on the un-filtered data
after the 130$^{th}$ $sec$ still largely remains. The same applies to
the transformation of panel 9b) of Figure
\ref{wavelet_lower_filtered}: All high values in very low
periodicities have disappeared up to 130$^{th}$ $sec$, but a
significant number remains after then. On higher periodicities there
is clearly an amplification of existing high values mainly on the
region after the 130$^{th}$ $sec$. The increase in power on the low
periodicities coincides with a similar increase in the power of high
periodicities.

\section{Significance test using Monte Carlo analysis}

The statistical analysis on SECIS 2001 as performed by K03 \& K04
indicates that there are \lq \lq false'' (i.e.\ caused by noise or
atmospheric effects) oscillations that satisfy the criteria establish
by K03. In particular atmospheric seeing is known to produce
differential distortion effects that produce \lq \lq false''
detections in the range of 5 to 19 $sec$ periods. So far the only
satisfactory way found to treat those false detections is
statistical. The possibility for a detection to be due to noise or
atmospheric effects was calculated by scanning for oscillations large
parts of the data were no \lq \lq real'' (i.e.\ solar coronal)
oscillations can be expected. The number of oscillations found in
those areas was then used to establish the possibility of detection to
be false in the areas of the lower solar corona that are in the
proximity of the \lq \lq test'' areas. The cases were the
concentration of detections is much higher than expected, a detection
of a corona wave is reported (as in K04).

In a bid to establish a quantitative criterion to determine which
detections of oscillations are due to noise or atmospheric effects and
overcome the limitations of the statistical methods used, the Monte
Carlo analysis (or randomisation) was investigated. This method has
successfully used before in the analysis of solar physics data by
\inlinecite{Oshea,Banerjee} who applied it to time series analysed by
wavelet transforms. It was chosen because it provides a test of noise
that is distribution free (or else non-parametric), i.e.\ it does not
depend on any given noise model (e.g.\ Gaussian noise, Poisson,
etc). Here we follow the Fisher method as described by 
\inlinecite{Nemec} and performed 1000 permutations per pixel of the 
aligned three-dimensional array. In order to evaluate the performance
of this randomisation method, we used the same test areas of the Moon
and the outer corona as in previous section. For each individual pixel
of these areas the maximum power of the wavelet transformation was
recorded and then compared to the maximum value of the 1000 shuffled
time series produced from the original. The percentage $p$ of those
shuffled time series that had maximum power of their wavelet
transformation larger than that of the original time series was then
recorded. When the original time series is random noise of any given
type (as this test is distribution free) we expect that 50\% of the
shuffled time series will have wavelet transformations with higher
maximum power than the original data. For the purposes of this
analysis we will consider any value $p$ smaller than 1\% (ie less than
1\% of the shuffled time series has wavelet transformation with power
more than that of the original time series) is indicating that the
original time series has a strong signal comparing to the noise level.

\begin{table}
\begin{tabular}{|c|c|c|}
\hline
Percentage of $p \geq 1\%$ (\%) & Before noise filtering & After
noise filtering\\
\hline
Moon                 & 60                     & 15      \\
Outer Corona         & 65                     & 10      \\
Lower Corona         &  5                     &  1      \\
\hline
\end{tabular}
\caption[]{Percentage of pixels in each of the test areas before and 
after the noise filtering that were found to have $p \geq 1\%$.}
\label{table}
\end{table}

Table \ref{table} contains the percentage of pixels in each of the
test areas that were found to have $p \geq 1\%$. A number of
important results become apparent: First of all the areas of the Moon
and outer corona before noise-filtering have more that the half of
their pixels having a relatively low S/N ratio, while the percentage
of the lower corona pixels (again before filtering for noise) that
have a similarly low S/N ratio is much lower (5 \%). This is
significant as it confirms the choice of $p \geq 1\%$ as a criterion
to distinguish between the pixels that have signal dominated by the
solar corona and those that do not. Second the application of the {\it
\`a Trous} noise filtering makes reduces the \lq \lq randomness'' of 
the data set significantly (by a factor of $\approx$ 5) making it
difficult to use the randomisation test reliably. A third useful
result shown on Table \ref{table} is that the percentage of pixels
that have $p \geq 1\%$ is approximately the same in the Moon area
(where we know that the signal of all pixels is due to not direct
observation of the Sun) and the outer corona (where we know the signal
is partially directly from the solar corona and partially scattered
light). This is important as it indicates that the scattered light on
this image area is a significant portion of the signal and it will be
very difficult to distinguish which detections are \lq \lq real'' in
this area by applying the existing criteria.

{\it \`A Trous} wavelet transformation's inability to reduce the
number of \lq \lq false'' detections can be explained if we
consider the different contributions to the signal. Any pixel value of
this data set is a combination of the detection of scattered light from
Earth's atmosphere, Gaussian noise and (for those areas that are not
covered by Moon) the detection of light directly from the solar
corona. By removing the Gaussian noise any weak oscillations due to
Earth's atmosphere could be detected more clearly. Those
oscillations are known to be introduced by various optical effects
produced by Earth's atmosphere: variation in transmission through the
atmosphere, differential distortions caused by winds at high altitude,
etc. In contrast the area lead by the lower solar corona had already
enough signal and a high S/N ratio, therefor the reduction of the
Gaussian noise levels did not contribute to a major increase in the
number of detected oscillations. Also because the signal coming from
the solar corona in this particular area was strong before the {\it
\`a Trous} noise filtering, the relatively weaker atmospheric
effects did not increase dramatically after the subtraction of the
Gaussian noise. As a result the number of detected oscillations did
not increase as dramatically in that area as in the other two.

The most significant breakthrough of the efforts to establish a
quantitative criterion to determine which detections of oscillations are
\lq \lq real'' and which are not, came when $p$ was calculated for the 
pixels that were found to oscillate by K04 (i.e.\ the 5 pixels
of the Moon area, 11 pixels of the outer corona area and 66 pixels of
the lower corona). All pixels from lower corona were found to have $p
< 0.1\%$ (i.e.\ none of the 1000 shuffled time series was found to
have a wavelet transformation with higher power than that of the
original time series), while 4 out of 5 of the detections from the
Moon area and 10 out of 11 detections from the outer corona, were
found with $p > 0.1\%$. Therefor, a criterion can be establish that
will use the randomisation test described here to reject those pixels
with $p> 0.1\%$.

\section{Discussion}

Two well known signal processing techniques, The \`a Trous wavelet
transformation and Monte Carlo analysis, were applied to SECIS 2001
data. Those two methods were evaluated by using two \lq \lq test''
areas (areas were the signal from the solar corona was expected to be
small or zero) and a \lq \lq useful'' area were detections of corona
oscillations were expected. By comparing the results from the three
areas, an accurate evaluation of the numerical techniques described
above was made.

The \`a Trous algorithm produced mixed results. Although the reduction
to Gaussian noise level was very significant, the ability to detect
corona waves was actually reduced. This is because of the effect of
Earth's atmosphere in the data set. Intensity oscillations caused by
the atmosphere were weaker in signal than those caused by the solar
corona, therefor, when the S/N ration was lower those oscillations did
not obtain the significance levels needed to become detections. After
the noise was reduced the significance levels of the intensity
oscillations caused by the atmosphere was increased enough to produce
a large number of false detections. It is also worth noticing that the
areas where the signal from the corona was stronger had much less
false detections than those areas with weak or no signal, indicating
that the signal from the lower corona is significantly stronger than
the atmospheric effects.

By the use of the two \lq \lq test'' areas of the data set it has
become apparent that a reliable, objective, numerical method is needed
in order to distinguish those detections caused by plasma from the
solar corona to those introduced by atmospheric effects. The Monte
Carlo analysis (otherwise refereed here as randomisation test) was
investigated as a means to make this distinction. All detections
reported as \lq \lq real'' by K04 were found to have $p < 0.1\%$ while
almost all (14 out of 16) of those reported as \lq \lq false'' were
found to be in the range of $100\% > p > 0.1\%$. Therefor the value of
$p$ is proposed as a criterion for rejecting future detections from
the lower corona region.

\acknowledgements
This work was completed under the ESA/PRODEX contract C90117 "SWAP
Preparation for Exploitation".

\end{article}

\begin{thebibliography}{}

\bibitem[\protect\citeauthoryear{Aschwanden}{2004}]{Aschwanden04}
Aschwanden M.J., 2004, {\it Physics of the Solar Corona, An
Introduction}, Springer-Verlag Telos

\bibitem[\protect\citeauthoryear{Banerjee}{2001}]{Banerjee}
Banerjee D., O'Shea E., Doyle J.G., Goossens M., 2001, A\&A, 380, L39

\bibitem[\protect\citeauthoryear{Holschneider et al.}{1989}]
{Holschneider} Holschneider M., Kronland-Martinet R., Morlet J.,
Tchamitchian P., 1989, A real-time algorithm for signal analysis with
the help of the wavelet transform, in {\it Wavelets: Time-Frequency
Methods and Phase-Space}, 286, Springer-Verlag

\bibitem[\protect\citeauthoryear{Katsiyannis et al.}{2003}]
{Katsiyannis03} Katsiyannis A.C., Williams D.R., McAteer R.T.J.,
Gallagher P.T., Keenan F.P., Murtagh F., 2003, A\&A, 406, 709

\bibitem[\protect\citeauthoryear{Katsiyannis et al.}{2004}]
{Katsiyannis04} Katsiyannis A.C., Williams D.R., Murtagh F.D.,
McAteer R.T.J., Keenan F.P., 2005, ESA SP-575, 410

\bibitem[\protect\citeauthoryear{Koutchmy et al.}{1983}]{Koutchmy83} 
Koutchmy S., \v{Z}ug\v{z}da Y.D., Loc\v{a}ns V., 1983, A\&A, 120, 185

\bibitem[\protect\citeauthoryear{Nakariakov}{2003}]{Nakariakov03}
Nakariakov V.M., 2003, in {\it Dynamic Sun}, Ed.~B.~Dwivedi, CUP

\bibitem[\protect\citeauthoryear{Nakariakov et al.}{2004}]
{Nakariakov04} Nakariakov V.M., Arber T.D., Ault C.E.,
Katsiyannis A.C., Williams D.R., 2004, MNRAS, 349, 705

\bibitem[\protect\citeauthoryear{Nemec \& Nemec}{1985}]{Nemec}
Nemec A.F. \& Nemec J.M., 1985, AJ, 90, 2317

\bibitem[\protect\citeauthoryear{O'Shea et al.}{2001}]{Oshea} 
O'Shea E., Banerjee D., Doyle J.G., Fleck B., Murtagh F., 2001, A\&A,
368, 1095

\bibitem[\protect\citeauthoryear{Pasachoff \& Landman}{1984}]
{Pasachoff84} Pasachoff J.M., Landman D.A., 1984, Sol.\ Phys., 90,
325

\bibitem[\protect\citeauthoryear{Pasachoff \& Ladd}{1987}]
{Pasachoff87} Pasachoff J.M., Ladd E.F, 1987, SP 109, 365

\bibitem[\protect\citeauthoryear{Phillips et al.}{2000}]{Phillips00}
Phillips K.J.H., Read P., Gallagher P.T., Keenan F.P., Rudawy P.,
Rompolt B., Berlicki A., Buczylko A., Diego F., Barnsley R., Smartt
R.N., Pasachoff J.M., Badcock B.A., 2000, Sol.\ Phys., 193, 259

\bibitem[\protect\citeauthoryear{Roberts et al.}{1984}]{Roberts84}
Roberts B., Edwin P.M., Benz A.O., 1984, ApJ, 279, 857

\bibitem[\protect\citeauthoryear{Starck \& Murtagh}{2002}]{Starck02}
Starck J.-L., Murtagh F., 2002, `Astronomical Image and Data Analysis', 
Springer-Verlag Berlin Heidelberg

\bibitem[\protect\citeauthoryear{Torrence \& Compo}{1998}]{Torrence98}
Torrence C., Compo, G.P., 1998, Bull. Amer. Meteor. Soc., 79, 61

\bibitem[\protect\citeauthoryear{Williams et al.}{2001}]{Williams01} 
Williams D.R., Phillips K.J.H., Rudawy P., Mathioudakis M., Gallagher
P.T., O'Shea E., Keenan F.P., Read P., Rompolt B., 2001, MNRAS, 326,
428

\bibitem[\protect\citeauthoryear{Williams et al.}{2002}]{Williams02}
Williams D.R., Mathioudakis M., Gallagher P.T., Phillips K.J.H.,
McAteer R.T.J., Rudawy P., Keenan F.P., Katsiyannis A.C., 2002, MNRAS,
336, 747

\end{thebibliography}
\end{document}